\documentstyle[twoside,fleqn,espcrc2]{article}

\newcommand{\beq}{\begin{equation}}
\newcommand{\eeq}{\end{equation}}
\newcommand{\beqa}{\begin{eqnarray}}
\newcommand{\eeqa}{\end{eqnarray}}

\title{Real Time Simulations in Lattice Gauge Theory}

\author{Guy D. Moore\address{Physics Department, McGill University,
	3600 University Street, Montreal, QC H3A 2T8 Canada}}
       
\begin{document}
\input epsf.tex

\begin{abstract}
I review the study of real (Minkowski) time correlators in hot, weakly
coupled Yang-Mills theory via lattice methods.  I concentrate on the
Minkowski time topological susceptibility, which is 
related to the efficiency of baryon number violation at high temperature.
It can be computed by approximating the IR fields as classical and
solving their dynamics nonperturbatively
on the lattice.  However it is essential to include the UV degrees of
freedom.  Their influence can be computed perturbatively and included
in the lattice model by the addition of auxiliary fields.
\end{abstract}

\maketitle

\section{Introduction}

Most of the work to date in lattice gauge theory has focused on
computing Euclidean quantities.  While this is a valid pursuit, 
there are very interesting questions to be asked about time-like
separated correlators.  For instance, almost all questions about
dynamics involve such correlators.  For QCD it
appears at this time that we have no first principles
nonperturbative tools to study these questions.  However,
the situation is somewhat different in the case of weakly
coupled, hot Yang-Mills theory.  This is a relevant question for the
SU(2) sector of the standard model, for which the coupling is weak,
$\alpha_w \simeq 1/30$.  Naively we would think that perturbation
theory should then work, and nonperturbative tools are not needed.
However at very high
temperatures this is not the case.  High
temperature questions are also interesting in cosmology; in particular
we may need to understand the behavior of the SU(2) weak sector at high
temperatures to understand the origin of the cosmological baryon
number abundance.

\section{Baryon number violation and topological susceptibility}

\begin{figure}[htb]
\vspace{0.9in}
\includegraphics{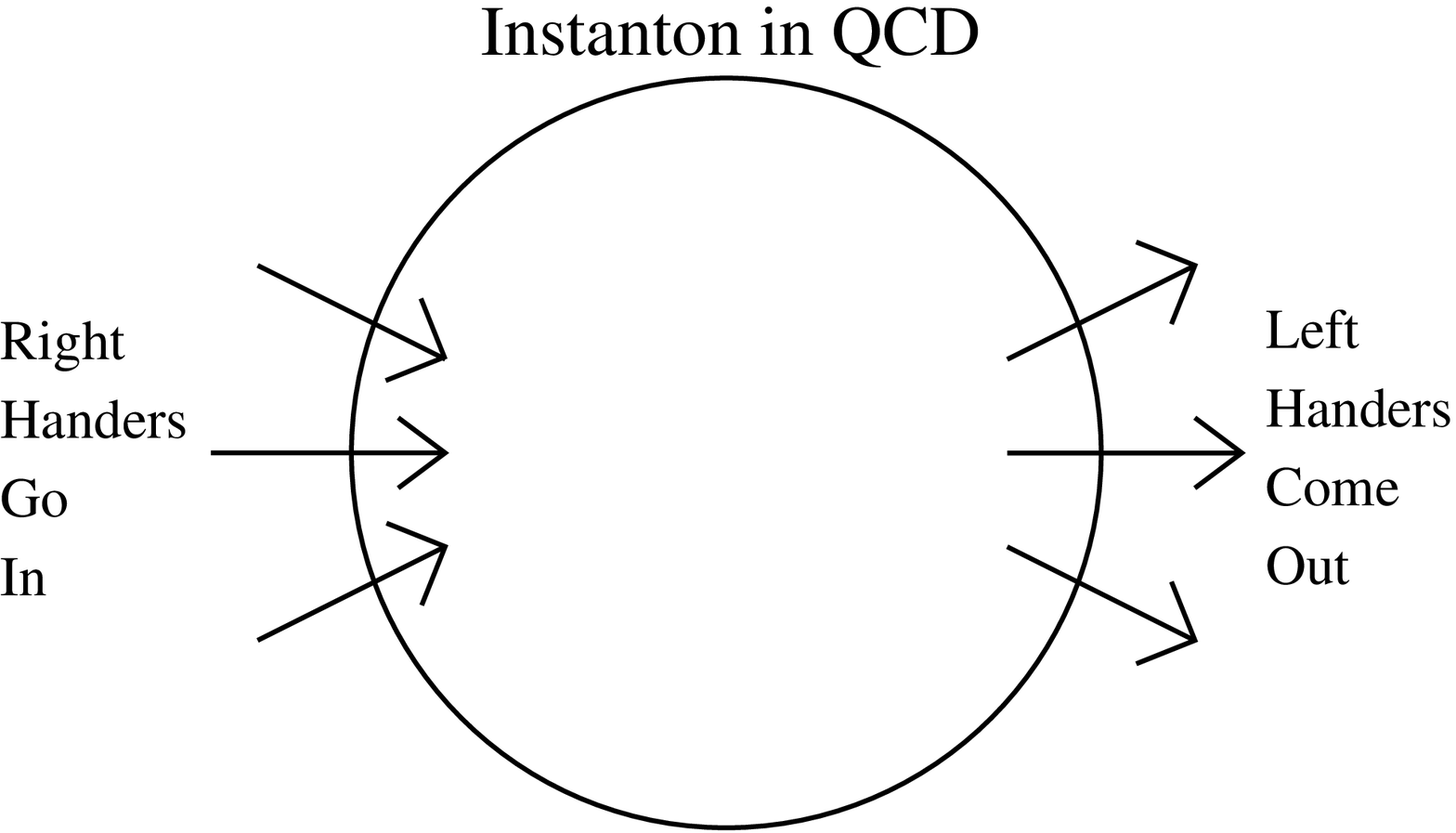}
\vspace{1.5in}
\includegraphics{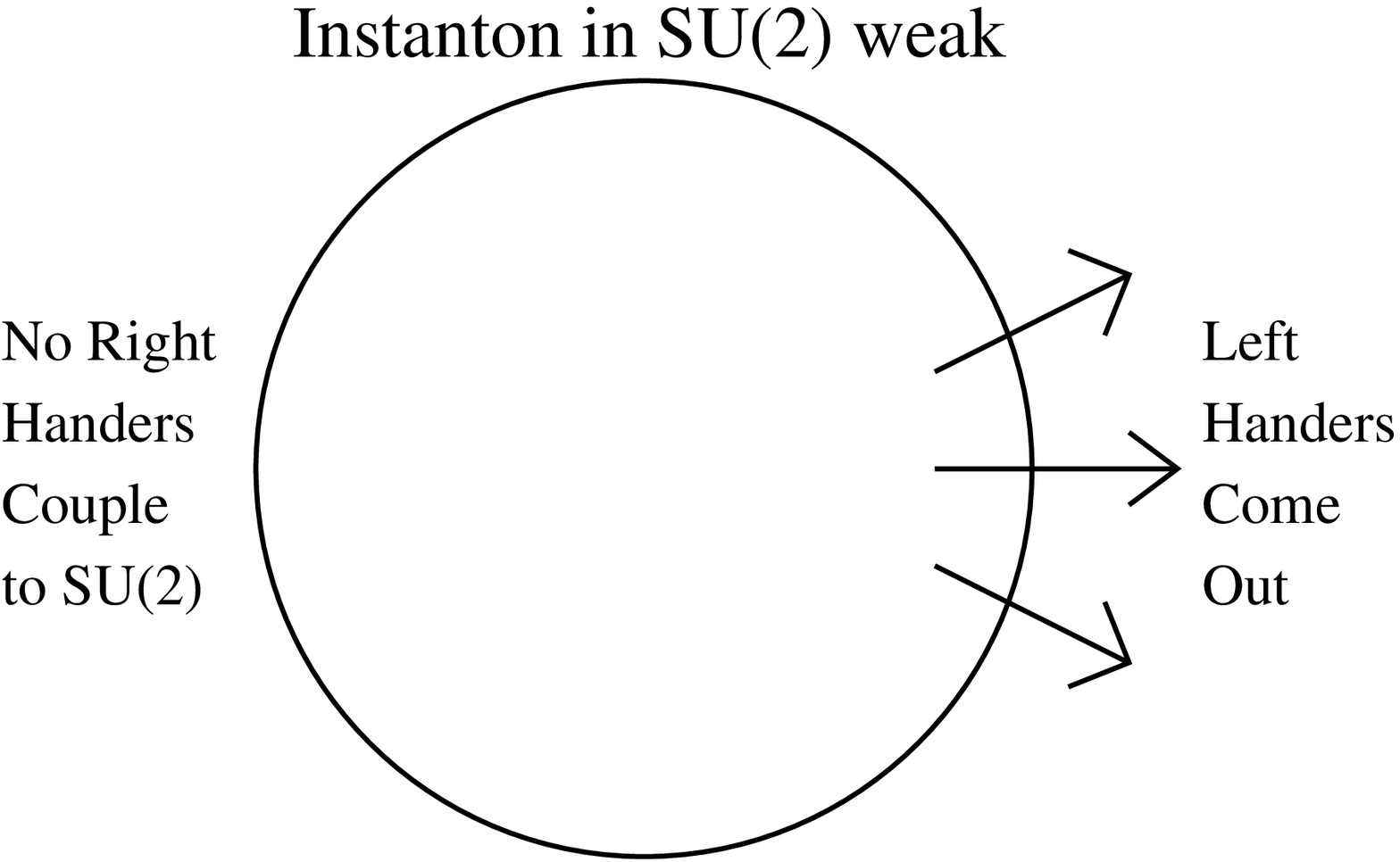}
\caption{\label{instanton}
Top:  cartoon of what an instanton in QCD does; it turns
right handed quarks into left handed ones.  But because SU(2) weak
only couples to left handed fermions, a weak instanton (Bottom) creates
a net particle number.} \vspace{-0.2in}

\end{figure}

First, a big question; how efficiently is baryon number
violated in the standard model?  To see that it is violated, recall
what happens when there is an instanton in QCD:  left
handed quarks go in, right handed quarks go out (see
Fig. \ref{instanton}).  

Instantons are also permitted in SU(2) weak; the difference is that only
left handed fermions couple to SU(2) weak, and so, while left handers
go out, no right handers go in.  This is particle number violation; in
particular, baryon number is violated, as first noted by t'Hooft
\cite{tHooft}.  

So why don't we all die?  Because the decay constant for baryons at
normal temperatures is tiny.  In vacuum we can relate the baryon
number violation rate to the density of instantons in the Euclidean
theory, giving a decay rate of order 
$\exp(-16 \pi^2 / g_w^2) < 10^{-160}$.

The situation may be different at high temperatures.  Arnold and
McLerran \cite{ArnoldMcLerran} have shown that the 
decay lifetime $\tau$ for a baryon number excess is related to
the Minkowski space topological susceptibility (often called the
sphaleron rate),
\beqa
\Gamma & \equiv & \int_{- \infty}^{\infty} dt \int d^3 x 
	\left( \frac{g_w^2}{32 \pi^2} \right)^2
	\nonumber \\ & & \qquad \times
	\langle F_{\mu \nu}^a \tilde{F}^a_{\mu \nu}(x,t)
	F_{\alpha \beta}^b \tilde{F}^b_{\alpha \beta}(0,0) \rangle \, ,
\label{Gamma_defined}
\eeqa
through $\tau^{-1} = (39/4 T^3) \Gamma$.  Here $\Gamma$ should be
computed using the thermal density matrix but at zero chemical
potential.  
This relation applies for temperatures $T >> m_{\rm proton}$ and
baryon number densities $N_{\rm B}/V \ll T^3$.

Note that $\Gamma$ is NOT the Euclidean topological
susceptibility;  Eq. (\ref{Gamma_defined}) involves an integral over
real (Minkowski) time.  The two susceptibilities may be related in
vacuum, but they are certainly not at high temperature;
the Euclidean susceptibility is always exponentially
small and falls with temperature, while $\Gamma$ grows with
increasing temperature, eventually becoming $O(\alpha_w^5 T^4)$.

To compute $\Gamma$ we need a nonperturbative way to measure 
$\int \langle F_{\mu \nu}^a \tilde{F}^a_{\mu \nu}(x,t)
F_{\alpha \beta}^b \tilde{F}^b_{\alpha \beta}(0,0) \rangle$ 
at large $T$.  Because the time integral involves all times, we cannot
do this reliably by analytic continuation from Euclidean time.
And as we shall now see, perturbation theory
also proves unreliable. 

\section{Thermal Perturbation Theory and Classical Physics}

My discussion here will follow \cite{Aarts}.
One can arrange thermal perturbation theory in the real time formalism
in terms of order symmetrized ($F$) and antisymmetrized (retarded and
advanced) propagators.  Each diagram gives several
contributions in which different propagators are $F$ or
advanced/retarded, and as we add loops to a diagram, 
the maximum possible number of $F$ propagators increases by 1 per loop.  In
this formalism, temperature appears as the replacement on $F$
propagators of the vacuum zero point fluctuation amplitude with the
thermal fluctuation amplitude; for bosons,
\beq
\frac{1}{2} \rightarrow \frac{1}{2} + \frac{1}{\exp(E/T)-1} 
	\, .
\eeq
Neglecting interactions $E=\sqrt{k^2+m^2}$.  At high temperatures the
Higgs condensate, responsible for particle masses, dissolves and $m
\simeq 0$.  In the infrared, the Bose distribution function is large
and we can make the expansion
\beq
\frac{1}{2} + \frac{1}{\exp(E/T)-1} = \frac{T}{E} + \frac{1}{12}
\frac{E}{T} - \frac{1}{720} \frac{E^3}{T^3} \ldots \, ,
\label{bose_expansion}
\eeq
which in the infrared is very well approximated by $T/E$.  The same
expansion for the fermions starts at $O(E/T)$, so their IR effects are
negligible.

If we start out with the classical theory we get the same
perturbation theory except that only diagrams with the maximum number
of $F$ propagators appear, with Eq. (\ref{bose_expansion}) simply
replaced by $T/E$.  Hence, the IR of a weakly coupled quantum theory at
$T \gg m$ behaves at leading order as a classical theory.

Moreover, this leads to a failure of perturbation theory.
In vacuum, the loop counting expansion parameter is $\alpha_w$, which
is small; but if one propagator per loop is an $F$ propagator,
the expansion parameter becomes $(T/E) \alpha_w$.  If some bosonic
mass scale is $m \leq \alpha_w T$, then for $k \sim \alpha_w T$, the
expansion parameter is $1$ and perturbation theory breaks down.
However,
perturbation theory only fails in the IR, and
it fails only where the fields behave classically up to
small corrections.
Therefore we can treat the strongly coupled IR
behavior by treating the IR degrees of freedom classically.

\section{Classical Theory on the Lattice}

With this in mind, Grigoriev and Rubakov conjectured that
any unequal time IR dominated correlator in hot Yang-Mills theory
would take the same value as its classical theory analogue, up to
$O(\alpha_w)$ corrections \cite{GrigRub}.  This turns out to be
incorrect; it is essential to include the interactions between the IR
modes and the (quantum but perturbative) UV modes \cite{ASY}.  
But for starters we will ``throw out'' the UV and just
look at $\Gamma$ in the regulated classical theory.

We need a nonperturbative regulation which will preserve exact gauge
invariance and remove the unwanted (nonclassical) UV degrees of
freedom; naturally the lattice is the best candidate.  Real time
classical Yang-Mills theory on the lattice looks almost like
the molecular dynamics algorithm applied to the 3-D path integral
without fermions.  To each link matrix $U_i \in {\rm SU(2)}$ we
associate a canonical momentum $E_i$ in the Lie algebra, satisfying
equations of motion \cite{Ambjornetal}
\vspace{-0.05in}
\beqa
\frac{dU_i}{dt} & = & i \tau^a E^a U_i \, , \\
\frac{dE_i}{dt} & = & - \sum_{\Box} \frac{1}{2} 
	{\rm Tr} i \tau^a U_{\Box} \, ,
\eeqa
\vspace{-0.07in}
which arise from the Hamiltonian
\beq
H = H_{\rm Wilson, \; 3-D} + \sum_{x,i} \frac{1}{2} E^a_i(x) E^a_i(x)
	\, .
\eeq
\vspace{-0.05in}
There are two differences from molecular dyamics.  First, the $E$
fields must satisfy Gauss' Law,
\vspace{-0.05in}
\beq
D_L \cdot E = 0 \, ,
\eeq
\vspace{-0.05in}
which is conserved by the equations of motion but must be enforced on
the initial conditions and considered in the thermalization.  Second,
we interpret $t$ not as a ``fake'' time put in to perform a
Monte-Carlo, but literally as Minkowski time.  

\begin{figure}[htb]
\vspace{1.2in}
\includegraphics{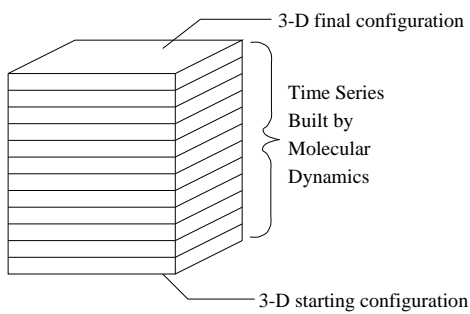}
\caption{\label{topology_pic} The 4-D space for which we must measure
toplogy.}
\end{figure}

To determine $\Gamma$ we must measure the topology of the 4-D
configuration gotten by taking 3-D configurations at successive times
as neighboring 3-D slices of a 4-D lattice, see
Fig. \ref{topology_pic}.  

\vspace{-0.2in}

\noindent
The space has open boundary conditions, so
topology is technically undefined.  Also we need a very
efficient algorithm, since the algorithm which generates the
configurations is very fast.  We fix the first problem
by repeatedly appending a ``new'' initial and final 3-D slice, equal
to the old one after some 3-D cooling.  Sufficient cooling leads
quickly to the vacuum (because we are in 3-D), and topology is defined
with vacuum boundaries.  The topological suceptibility is not affected
in the limit of large molecular dynamics time.  
Finding an efficient
algorithm is made easier because the gauge configurations are very
smooth; first, the mean plaquette is much closer to ${\bf 1}$
than in the QCD setting because of the difference between 3-D and 4-D
thermodynamics (not surprising as the underlying coupling is
weak).  Also, the temporal spacing is very small.  Cooling
within 3-D planes makes topological configurations spread in the 3
dimensions; they contract in the temporal direction, but it is highly
resolved anyway and this is not a problem.  It then remains to
integrate $F \tilde{F}$ with an $O(a^2)$ improved operator.  
The details are in \cite{broken_nonpert}.

\begin{figure}[htb]
\vspace{2.0in}
\includegraphics{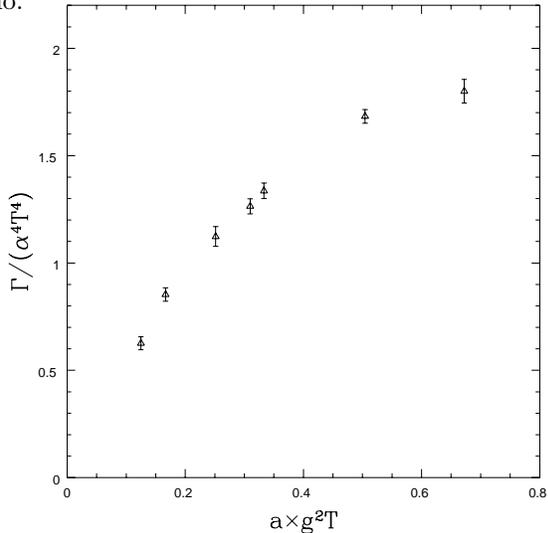}
\caption{\label{YM_data} Topological suceptibility of hot lattice
SU(2) Yang-Mills theory plotted against lattice spacing.  Linear
lattice spacing dependence is evident at the smallest lattice
spacings.}
\end{figure}

Is $\Gamma$ in the classical theory lattice spacing independent?  The
data \cite{MooreRummukainen}, presented in Fig. \ref{YM_data}, says
no. 

\vspace{-0.15in}

This behavior is expected analytically \cite{ASY}; it means that the
UV must matter, and we should treat it properly.  

\section{Putting in the UV physics}

We need to include properly the interactions between IR and UV.
The UV does not behave classically;
but it behaves perturbatively, and its leading order 
influence on the IR, the Hard Thermal Loops (HTL's), is known
\cite{HTL}.  The
HTL effective action is nonlocal \cite{HTLaction}, but it is
equivalent to the effect of a large collection of classical, adjoint
charged particles \cite{Kelley}.  This means we may be able
to incorporate the effects in a local way, and it provides the
intuition of why we have to.

First, the intuition.  The UV modes are short wavelength and so behave
like classical particles.  A
large collection of charged particles is a plasma, and plasmas are
conducting.  By Lenz's Law, conducting media resist changes in magnetic
fields.  But evolution of IR magnetic fields is precisely what is
required to get topology change.  Hence, the more UV modes there are,
the smaller $\Gamma$ should be.  This is basis of the
arguments of Arnold, Son, and Yaffe \cite{ASY,ASY2}.

How should we add the UV effects to the lattice IR theory?  Two
approaches seem reasonable; an N-body treatment and a
Boltzmann-Vlasov treatment.

\begin{figure}[htb]
\vspace{1.5in}
\includegraphics{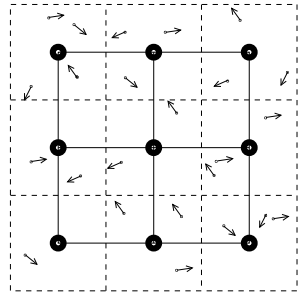}
\caption{\label{particle_cartoon}Cartoon of how ``particle'' degrees
of freedom live in the continuous space between lattice sites.  The
charge resides at the vertex nearest the particle.}
\vspace{-0.18in}
\end{figure}

In the N-body approach, we add a large number of ``particle'' degrees
of freedom to the lattice gauge theory \cite{HuMuller}.  Each particle
has a coordinate, a momentum, and an adjoint representation charge.
In Hu and M\"{u}ller's formulation, the coordinates and momenta are
each continuous degrees of freedom, so the particles ``live'' in the
continuous space between the lattice sites, see
Fig. \ref{particle_cartoon}.  

The exact rules for the interactions between the particles and the
lattice degrees of freedom were worked out in \cite{particles}.  Each
particle charge resides at the nearest lattice point, and the particle
interacts with the lattice fields at the discrete instants when it
passes from being nearest one site to being nearest another.  The system has
a conserved energy and phase space measure and exactly respects Gauss'
Law; and in the small $a$ and small particle charge limit it correctly
reproduces the classical field + HTL effective theory.

The details of the Boltzmann-Vlasov approach appear in Kari
Rummukainen's contribution to these proceedings \cite{kari}, but I
briefly review them here.  Rather  than considering particles
individually, one considers continuous particle population functions
$f(\vec{p},\vec{x},q)$.  Linearizing in the departure from
equilibrium, $f \rightarrow \delta f^a(\vec{p},\vec{x})$, which in the
current setting can be further simplified by integrating over $|p|$,
leaving only its direction $\hat{p} \equiv \vec{v}$.  The
resulting population function is called $W^a(x,\vec{v})$, and up to
normalization it represents the net charge of particles moving in the
$\vec{v}$ direction at position $x$.  The Yang-Mills-Boltzmann
equations in continuum are \cite{BlaizotIancu}
\vspace{-0.05in}
\beqa
(D_\nu F^{\nu \mu})^a & \! \! = \! \! & j^\mu_a \, , \\
j_a^\mu & \! \! = \! \! & m_D^2 \! \int \! \frac{d\Omega_v}{4 \pi} 
	v^\mu W^a(x,\vec{v}) \, , \\
(v_\mu D^\mu)^{ab} W^b(x,\vec{v}) & \! \! = \! \! & v_\mu F^{0\mu}_a(x) \, ,
\eeqa
\vspace{-0.05in}
where $v^\mu = (1,\vec{v})$ and $d\Omega_v$ is the solid angle measure
for integrating over $\vec{v}$.

The continuum equations not only need to be regulated on a lattice,
discretizing space; they also need regulation on the sphere, which we
do by expanding in spherical harmonics and truncating the series
at a finite $l_{\rm max}$.  For even $l_{\rm
max}$ there is surprisingly little truncation error \cite{kari}.

\begin{figure}[htb]
\vspace{2.1in}
\includegraphics{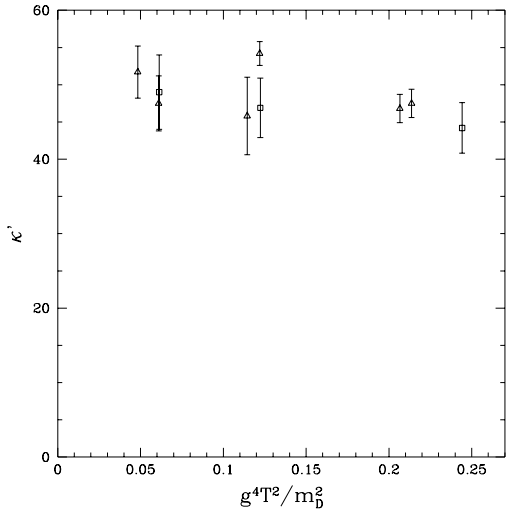}
\caption{\label{resultfig} $\Gamma$, scaled by the expected parametric
behavior, plotted against reciprocal conductivity.  Theory predicts a
flat line.  Data with squares are using the N-body approach, triangles
are with the Boltzmann-Vlasov approach.}
\end{figure}

In either approach, we expect the final answer to vary inversely with
the conductivity, which is proportional to the Debye mass squared
$m_D^2$.  It makes sense, then, to fit $\Gamma$ to the form
\beq
\Gamma = \kappa' \left( \frac{g^2 T^2}{m_D^2} \right) 
	\alpha_w^5 T^4 \, ,
\eeq
with $\kappa'$ to be determined.  The factor of $(\alpha T)^4$ is the
four powers of the nonperturbative scale $\alpha T$ required by
dimensions and the last power of $\alpha$ 
is the effect of the conductivity.
The numerical data from the two
approaches to the UV physics are in good agreement, as shown in
Figure \ref{resultfig}.

\vspace{-0.1in}

The result is $\kappa' \simeq 45$, which is abundantly fast enough to
erase any GUT generated initial baryon number (unless the GUT violates
$B-L$) and also fast enough to make electroweak baryogenesis viable if
there is sufficient CP violation and loss of equilibrium at the
electroweak phase transition.

\section{conclusion}

Real time techniques are available when the coupling is small.
Although naively perturbation theory should be reliable in this case,
it breaks down whenever some particle mass comes on order $m \sim
\alpha T$; but in this case the nonperturbative physics occurs only in
the infrared, which is classical and can be treated nonperturbatively
in real time on the lattice.

The influence of UV modes on the IR is very important in a gauge
theory, and there are now two different techniques, an N-body
approach and a Boltzmann-Vlasov approach, to include it.  The
techniques are in good agreement.

The Minkowski topological suceptibility of pure Yang-Mills theory has
been measured accurately, and this question is now closed.  However
there are still very interesting questions in nonperturbative IR
physics at the electroweak phase transition which remain to be
addressed, such as the dynamics of (Higgs) scalar condensates during
the phase transition.

\end{document}